# Hydrostatic Pressure Studies on Parent Phase SrFBiS$_2$ of BiS$_2$-based Superconducting Family

*Rajveer Jha, Brajesh Tiwari, and V.P.S. Awana*[*]
*CSIR-National Physical Laboratory, New Delhi-110012, India*

Recently discovered BiS$_2$-based layered superconductors have gained tremendous interest in the scientific community. These BiS$_2$-based superconductors have superconducting transition temperature (T$_c$) as 4.5K for Bi$_4$O$_4$S$_3$ [1,2], 2-5K for REO/FBiS$_2$ (La, Ce, Pr, Nd) [3-6] and 2.8K for Sr$_{0.5}$La$_{0.5}$FBiS$_2$ [7]. It is interesting to note that their parent phases LaOBiS$_2$ [3] and SrFBiS$_2$ [8] are semiconducting, which upon the O/F doping in former [3-6] and Sr/RE doping in later [7] become superconducting. Superconducting transition temperature in REO$_{0.5}$F$_{0.5}$BiS$_2$ RE (La, Ce, Pr, Nd) is observed to increase under the hydrostatic pressure by three fold just in 2GPa pressure [9-14]. It is suggested that this dramatic increase in T$_c$ is related to a structural phase transition under pressure especially for LaO$_{0.5}$F$_{0.5}$BiS$_2$ as reported by Tomita et.al [13]. Recently, we found that the T$_c$ of Sr$_{0.5}$La$_{0.5}$FBiS$_2$ superconductor increased by fivefold from around 2K to above 10K, accompanied by the semiconducting to metallic normal state [15]. Keeping in view, that application of hydrostatic pressure has resulted in tremendous increase of T$_c$ for the BiS$_2$ based superconductors [9-15], one wonders how their non-superconducting parent compounds will respond to the same. In the present work, we measure the temperature dependent electrical resistivity from 300K down to 2K under applied hydrostatic pressure (0-2.5GPa) for SrFBiS$_2$, which is the parent compound for the BiS$_2$–based superconductors [8].

In this study, the polycrystalline SrFBiS$_2$ sample was synthesized by the standard solid state reaction route via vacuum encapsulation. Stoichiometric ratio of Sr, SrF$_2$, Bi, and S were ground thoroughly in Ar-controlled glove box before palletized. Rectangular pellets were vacuum sealed (10$^{-3}$ Torr) in quartz ampoules and heat treated at 780$^0$C for 12h at heating rate of 2$^o$C/min with an intermediate grinding. X-ray diffraction (*XRD*) was performed at room temperature in the scattering angular (*2θ*) range of 10$^o$-80$^o$ in equal *2θ* step of 0.02$^o$ using *Rigaku Diffractometer* with *Cu K$_α$* (*λ* = 1.54Å). Rietveld analysis was performed using the standard *FullProf* program. The obtained phase pure samples were used for pressure dependent resistivity measurements by Physical Property Measurements System (*PPMS*-14T, *Quantum Design*) where pressure applied by HPC-33 Piston type pressure cell with Quantum design DC resistivity Option. Hydrostatic pressures were generated by a BeCu/NiCrAl clamped piston-cylinder cell. The sample was immersed in a fluid (Daphne Oil) pressure transmitting medium in a Teflon cell. Annealed Pt wires were affixed to gold-sputtered contact surfaces on each sample with silver epoxy in a standard four-wire configuration.

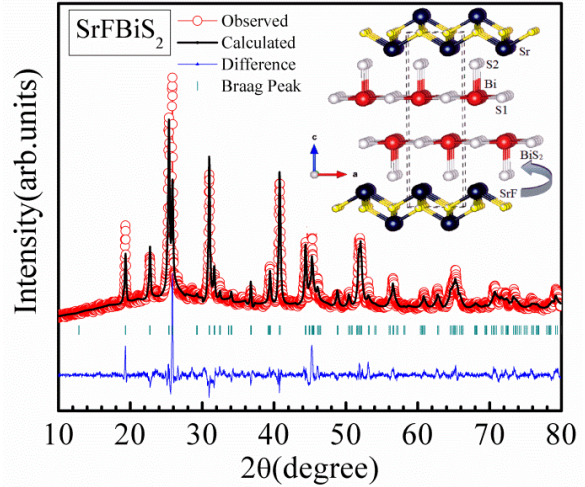

Figure 1 (Color online). Room temperature Reitveld fitted XRD pattern of SrFBiS$_2$

The room temperature observed and Reitveld fitted XRD pattern of SrFBiS$_2$ compound is shown in the figure 1. The SrFBiS$_2$ crystallized in tetragonal structure with space group *P4/nmm*. The estimated parameters are *a*=4.08(3)Å and c=13.78(2)Å as obtained from minimizing difference between the observed and calculated patterns, which are in confirmation with earlier report [8]. Inset of the figure 1 is the unit cell of the SrFBiS$_2$ compound.

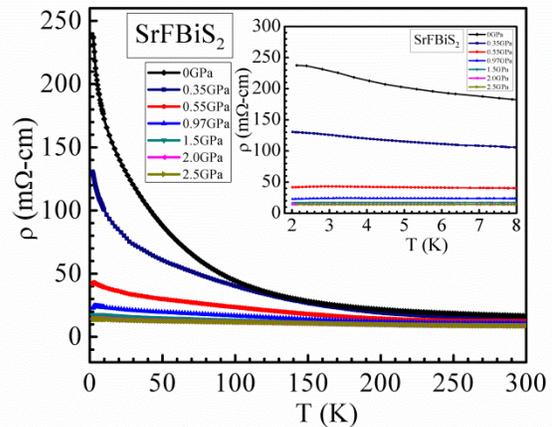

Figure 2 (Color online) Temperature (T) dependences of resistivity (ρ) under hydrostatic pressure for SrFBiS$_2$ and the inset is the extended part of the same from 2K-8K.

Figure 2 shows the temperature dependence of electrical resistivity under the different hydrostatic pressure (0-2.5GPa) for SrFBiS$_2$ down to 2K. A clear semiconducting behavior can be observed in the temperature range 300-2K as temperature coefficient

d$\rho$/dT<0 is negative. With increasing applied pressure the electrical resistivity decreases significantly. At the applied pressures of 2.17 and 2.5GPa the electrical resistivity becomes almost temperature independent. The inset of the figure 2 shows the zoomed portion of same in the temperature range 8K to 2K. There is no indication of superconducting transition down to 2K under maximum applied pressure of 2.5GPa. Though the suppression of resistivity in SrFBiS$_2$ under hydrostatic pressure is similar to the as for other BiS$_2$-based layered superconductors viz., LaO$_{0.5}$F$_{0.5}$BiS$_2$ and Sr$_{0.5}$La$_{0.5}$FBiS$_2$, but did not show superconductivity down to 2K under applied pressure of 2.5GPa. The absolute $\rho$ value at 2K is two orders of magnitude less than as in ref. 7 and even four orders lower than as in ref.8. This may happen due to the volatility of Bi, S and F in these compounds [7,8], and is in fact debatable.

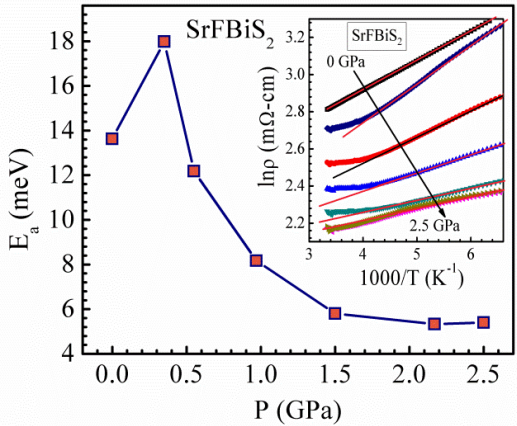

Figure 3 (Color online) Pressure dependence of activation energy (Ea), being evaluated from thermal activation equation; the solid line is a guide to eye. Inset is the ln$\rho$ Vs 1/T for the different pressures (0-2.5GPa) for SrFBiS$_2$; the solid lines are the extrapolation to the fitted portion.

Figure 3 presents the thermal activation energy (E$_a$) as a function of applied hydrostatic pressure as obtained from fitting of the resistivity data to the equation $\rho(T)=\rho_0 \exp(E_a/k_B T)$ above 100K. Except for 0.35GPa, the thermal activation energy drops exponentially with pressure. Inset of the figure 3 is the plot of ln($\rho$) Vs 1/T to evaluate the activation energy of SrFBiS$_2$ from 0-2.5GPa pressure. It is interesting to note that the value of $\rho_0$ under different hydrostatic pressures remains almost same close to 7.2m$\Omega$-cm. This observation suggests that at highest temperature the resistivity of sample is almost constant even under different hydrostatic pressures. The activation energy decreases with increasing pressure and saturates close to 5.3meV. This suggests that the Fermi energy moves close to conduction band by increasing pressure, while the highest temperature resistivity remains constant. Detailed spectroscopic and theoretical investigations with and without pressure are warranted to validate this point on recently discovered SrFBiS$_2$ compound.

In summary, the temperature dependence of resistivity suggests that the activation energy for one of the parent compound of BiS$_2$-based superconductor SrFBiS$_2$ decreases under applied pressure. Further, superconductivity could not be induced down to 2K under applied pressure of as high as 2.5GPa.

**Acknowledgement:** Authors would like to thank their Director NPL India Prof. R.C. Budhani for encouragement. This work is financially supported by *DAE-SRC* scheme on search for new superconductors. R. Jha acknowledges the *CSIR* for the senior research fellowship.